\documentclass{aa}
\usepackage{graphicx}
\begin{document}
\headnote{Research Note}
\title{The effects of blending on the light curve shape of Cepheids}
\author{E. Antonello}
\institute{Osservatorio Astronomico di Brera, Via E.~Bianchi 46,
       I--23807 Merate, Italy}
\offprints{E. Antonello \\ \email{elio@merate.mi.astro.it}}
\date{ Received date; accepted date }
\titlerunning{Blending effects}
\authorrunning{E. Antonello}

\abstract
{A short analysis is presented of the effects 
on the cepheid light curve shape, i.e. on the Fourier parameters usually 
adopted for its description, of the blending of the stellar image with other
close stars. The conclusion is that, within reasonable error, the effects 
are in general small and the Fourier decomposition is confirmed to be a 
useful tool for pulsation mode discrimination. A large effect
has been found on the phase differences in a narrow period
range corresponding to the known resonance centers between pulsation modes.
}

\maketitle

\section{Introduction}
Cepheids are primary distance indicators for external galaxies and those
used for this application pulsate in the fundamental mode. First 
overtone mode Cepheids are brighter by about 0.4 mag than fundamental mode 
pulsators with the same period. Since the period--luminosity relation has an 
intrinsic dispersion, which depends on several parameters (e.g. different 
effective temperature or color, different reddening, contribution from stellar 
companions), it is essential to remove the contaminating stars that 
are pulsating in a different mode. The large surveys of the Magellanic Clouds 
performed by MACHO (e.g. Welch et al. \cite{we}), EROS (e.g. Beaulieu et 
al. \cite{bea}) and OGLE (e.g. Udalski et al. \cite{uda4}) projects 
proved that the Fourier decomposition is a good technique for discriminating 
the mode among short period ($P \la 6$ d) Cepheids. More recently, 
the technique began to be applied to Cepheids of farther galaxies in the 
Local Group, such as IC 1613 (e.g. Antonello et al. \cite{pa2}; Dolphin 
et al. \cite{dol}) and M33 (Mochejska et al. \cite{moc1}).

The large surveys offered also the opportunity of discussing the problems 
related to blending. Mochejska et al. (\cite{moc}) define the blending
as the close projected association of a Cepheid with one or more 
intrinsically luminous stars, which cannot be detected within the observed 
point-spread function by the photometric analysis. There
is some debate about the implications for the distance determination related 
to the blending and more generally to poor resolution of the stellar images in 
these galaxies. 
The blending also has other effects on the light and 
the color curves. Mochejska et al. (\cite{moc}) note that in 
the case of a red or blue companion the light curve exhibits a flatter
minimum. As regards binaries, it is well-known that the 
observed amplitude of the light curve is affected by the luminosity of a 
bright companion. Could it be that the blending, apart from producing a
lower amplitude, also mimics a different pulsation mode? 
Recently, we recalled 
that in principle such an effect on the Fourier parameters is 
small in the context of mode identification (Antonello et al. \cite{afm}). 
Here we report the results of simulations that support this conclusion,
and we discuss some unexpected characteristics.

\begin{figure}
 \resizebox{\hsize}{!}{\includegraphics{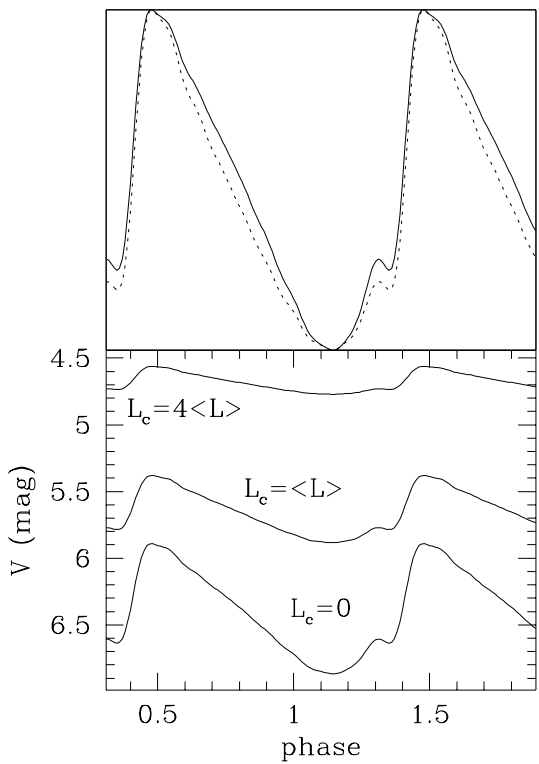}}
 \caption[ ]{Lower panel: blending effect on the $V$ light curve of a Cepheid,
for different values of the luminosity of the companion star ($L_c$). Upper
panel: comparison between the light curve for $L_c=0$ (continuous line)
and $L_c=4<L>$ (dotted line) scaled to the same amplitude.
} 
 \label{fit}
 \end{figure}

\section{Analysis}
The problem by itself would not be important if we adopt intensities 
instead of magnitudes to measure stellar brightness. Indeed, an increased 
intensity due to a close star, assuming no measurement error, would produce
a light curve with a similar shape to that without such a close star. The 
average intensity would be larger, the absolute amplitude would be the same,
and the relative amplitude would be of course decreased. Let $<L>$ be the 
average stellar intensity (that is, the average number of collected photons), 
$\Delta{L}$ the absolute amplitude, $A=\Delta{L}/<L>$ the relative amplitude, 
and $\epsilon \sim \sqrt{<L>}$ the mean absolute error on the measurement. 
Let us assume a close constant star with intensity $a<L>$. 
The relative amplitude of the system 
will be $A_1=\Delta{L}/[(a+1)<L>]$ and the mean absolute error 
$\epsilon_1 \sim \sqrt{(a+1)<L>}$. 
A close star has the effect of 
decreasing the relative amplitude and increasing the absolute error. 
This implies a lower order of fit of the reliable
Fourier decomposition of the intensity curve, and larger formal errors of 
the Fourier parameters; however, the parameters themselves are unchanged 
(within the formal errors).

The nonlinearity of the relation between intensity and magnitude introduces 
some changes. The simplest method for studying them is by means of
simulations. We considered light curves of some stars pulsating in the
fundamental or first overtone mode (e.g. X Cyg, DT Cyg)
observed by Moffett \& Barnes (\cite{mb}; data retrieved from McMaster Cepheid
Photometry and Radial Velocity Data Archive), and we adopted the best
fitting curve as a synthetic light curve. We simulated several time series,
adopting the original observing dates, and changing the synthetic light curve 
by introducing the contribution of a close constant star, and different mean 
errors of the measurement. In Fig. \ref{fit} we show the effects
of increasing luminosity on the synthetic light curve
of X Cyg. In the upper panel one can see the changes of light curve shape 
due to a four times brighter companion; the two curves are scaled to 
the same amplitude. The flattening of the minimum does not appear 
very prominent, even in this case where the magnitude difference 
between the Cepheid and the blended image is large, 1.75 mag. 

\begin{figure}
 \resizebox{\hsize}{!}{\includegraphics{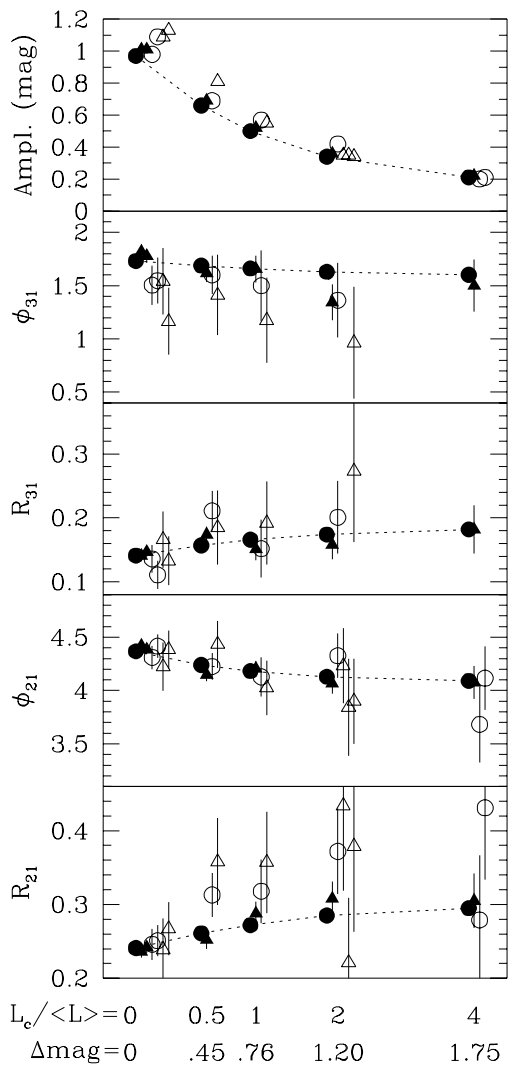}}
 \caption[ ]{The plots show how the Fourier parameters and light curve 
amplitude of a Cepheid change according to the luminosity of a companion star
($L_c/<L>$ is the ratio of the luminosity of the companion to the average
value of the Cepheid). The symbols indicate different values of the 
mean error $\sigma$ of measurements adopted in the simulations: 
{\em filled circle:} $\sigma$=0, {\em filled triangle:} 
$\sigma$=0.02; {\em open circle:} $\sigma$=0.05; 
{\em open triangle:} $\sigma$=0.1 mag. The errorbar indicates the formal
error of the respective parameter. $\Delta{mag}$ is the average magnitude 
difference between the Cepheid and the blended image
} 
 \label{four}
 \end{figure}

The time series were constructed applying a random number generator
for a Gaussian error distribution. The series were then Fourier decomposed
and the resulting Fourier parameters are plotted in Fig. \ref{four} for
the case of X Cyg, as an example. One can see clearly that the increasing 
blending implies a decreasing order of the reliable fit.

When performing the simulations, we also analyzed some OGLE stars
in the SMC, and we noted different trends with respect to the
above Cepheids. We suspected some dependence on the $P$, therefore we 
decided to analyze all the Cepheids in OGLE database of the SMC 
(Udalski et al. \cite{uda4}). The fitting curves of the 
Fourier decomposed $I$-band light curves were modified by introducing the
contribution of a companion star with $L_c=2<L>$, then they were analyzed
and we computed the difference between the Fourier parameters for
$L_c=2<L>$ and $L_c=0$. The results for the lowest order are shown in
Fig. \ref{diffe1} and \ref{diffe2} for the fundamental and first overtone mode,
respectively.
Although the effect on the amplitude ratio is always small, the trend with
$P$ is confirmed. The unexpected result is the large effect on the
phase difference very close to the resonance centers at $P \sim 10$ d for the
fundamental mode, and $P \sim 2.2$ d for the first overtone mode.
Outside these narrow $P$ ranges the effect is small.

\section{Discussion and conclusion}
\begin{figure}
 \resizebox{\hsize}{!}{\includegraphics{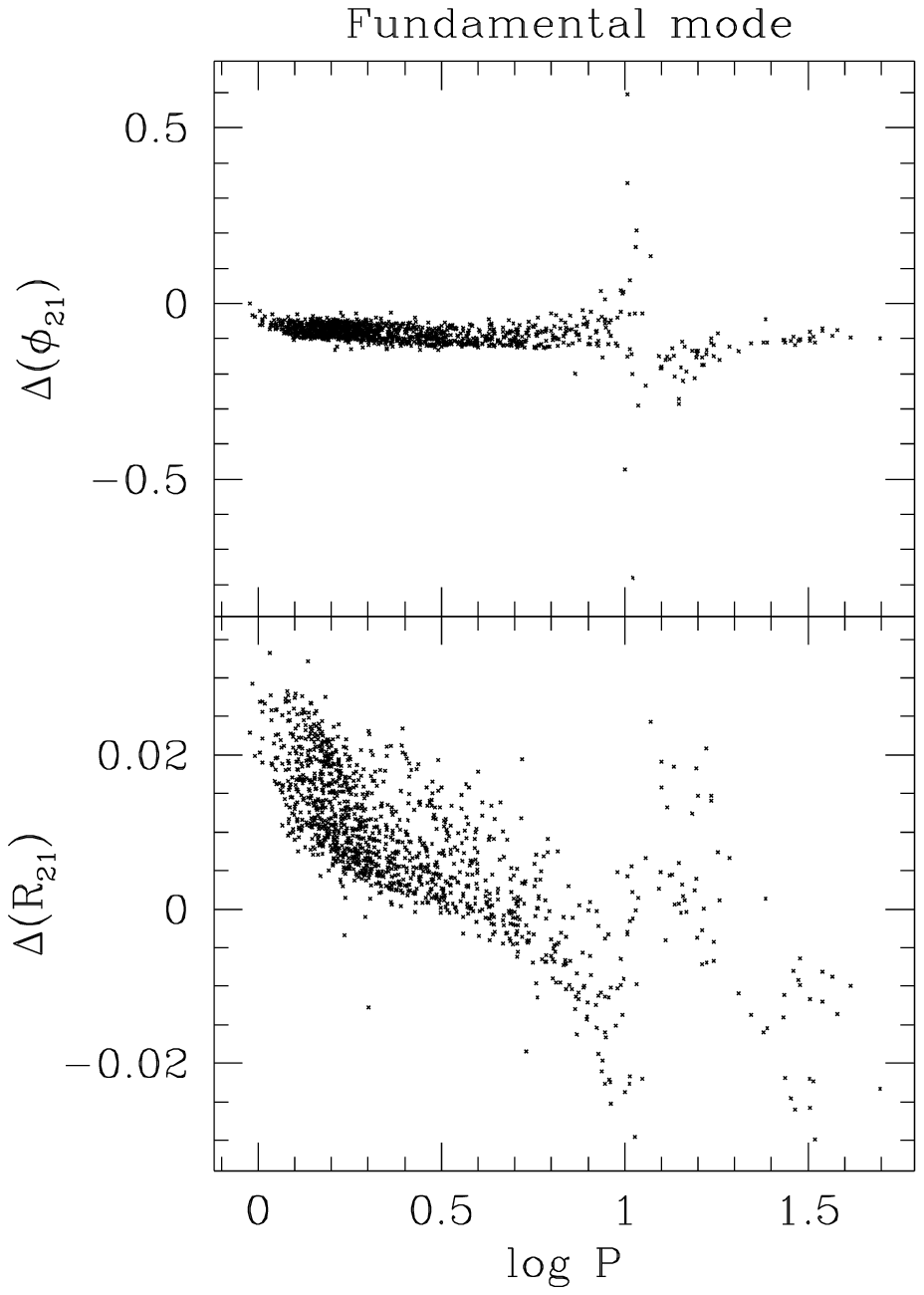}}
 \caption[ ]{Simulated blending effect on the $I$-band light curves of all the
OGLE fundamental mode Cepheids in the SMC.
The plots show the difference of $R_{21}$ and
$\phi_{21}$ between the light curves for 
$L_c=2<L>$ and $L_c=0$
} 
 \label{diffe1}
 \end{figure}
\begin{figure}
 \resizebox{\hsize}{!}{\includegraphics{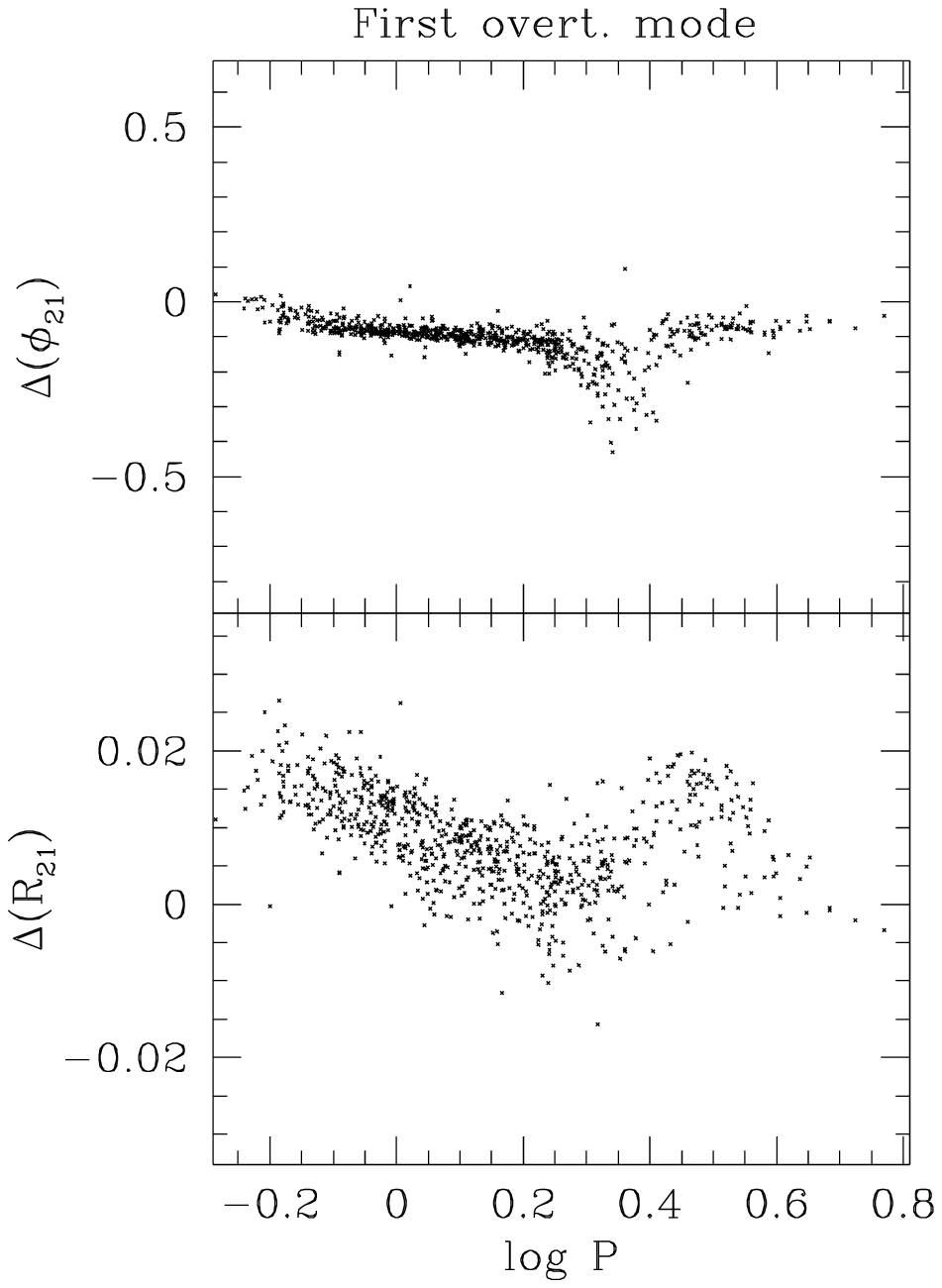}}
 \caption[ ]{Simulated blending effect on the $I$-band light curves of all the
OGLE first overtone mode Cepheids in the SMC.
The plots show the difference of $R_{21}$ and
$\phi_{21}$ between the light curves for 
$L_c=2<L>$ and $L_c=0$
} 
 \label{diffe2}
 \end{figure}
\begin{figure}
 \resizebox{\hsize}{!}{\includegraphics{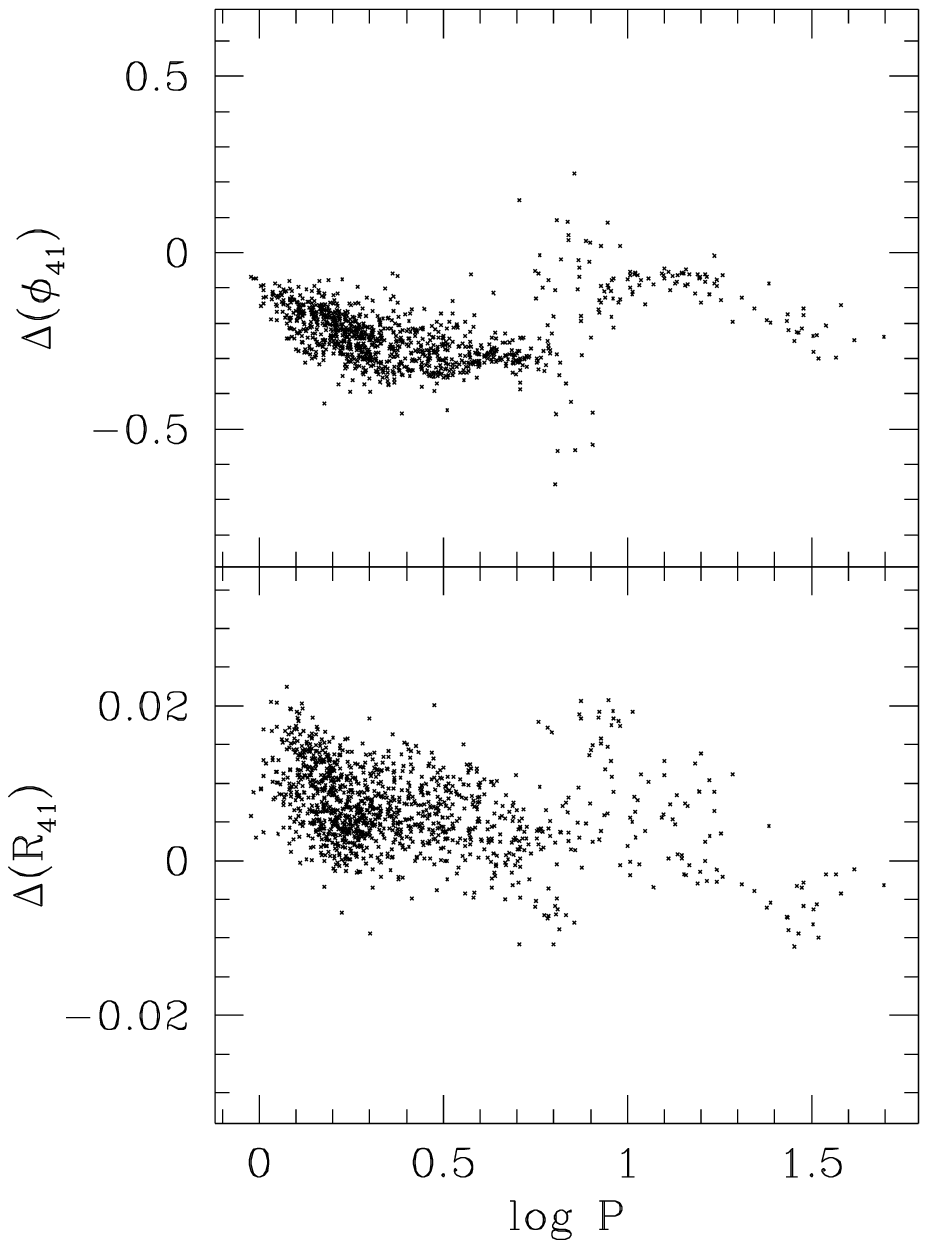}}
 \caption[ ]{
The difference of $R_{41}$ and $\phi_{41}$ between the simulated light curves 
for $L_c=2<L>$ and $L_c=0$ of fundamental mode Cepheids. It should be compared
with Fig. \ref{diffe1}
} 
 \label{diffe3}
 \end{figure}
The cases discussed here concern reasonable light curves; we do not
consider the problems related to very faint variables, which can hardly be 
detected at minimum light.
The requirement is that in the $P$ interval where it is possible
to find stars pulsating in different modes, the Fourier parameters must
allow us to make the discrimination. It is known that this occurs for 
$P \la 6$ d for the fundamental and the first overtone mode, using only 
light curves parameters. The results of the simulations show that in this 
$P$ range the blending has a negligible effect when we compare the differences 
introduced by it with the size of the parameters themselves. In particular, a 
blended fundamental mode pulsator will have slightly larger amplitude ratios 
than a non-blended one; we recall that the amplitude ratios of fundamental 
mode pulsators are intrinsically larger than those of first overtone mode 
ones in this $P$ range. The same occurs for a first overtone mode pulsator 
compared with a second overtone one, for $P \la 1.3$ d. On the other hand, 
a heavily blended first overtone pulsator increases its $R_{21}$ value, but 
in general not so much so as to be confused with a fundamental mode
pulsator. In conclusion, the blending due to various reasons is not an 
issue for the pulsation mode discrimination.

The color of the companion stars is not relevant for the present discussion, as
long as their contribution is constant; some (second order) effects could be 
related to their intrinsic variability, both in terms of photometric
variability and/or Doppler shift.
The influence of the photometric variability of the companion itself can be 
usually accurately estimated, since an adequate time series analysis is 
sufficient to disentangle the different contributions, because of the 
different periodicities or timescales involved. Also in this case,
however, it is wise to work with intensities rather than with magnitudes.
Variable seeing conditions could have some effect 
on the estimate of the intensity through the PSF fitting procedure; however 
in this case we would expect just an increased error in the measurement.

The plots in Figs. \ref{diffe1} and \ref{diffe2} suggest some interesting 
considerations. A light curve with an altered value of the mean luminosity,
such as that depicted in Fig. 1, or expressed with a different, nonlinear
mathematical function (e.g. the intensity instead of the magnitude) 
is characterized of course by (usually slightly) different Fourier parameters. 
If we estimate the differences related to these changes, we note that the
largest ones are for the phases of the Fourier components with smaller
amplitude; for example, at about 10 d some stars have $R_{21} < R_{i1}$, for
$i$ from 3 up to 6 or more. The large differences are not due to errors or to
uncertainties, since here we are not dealing with observed data but with 
synthetic light curves (i.e. the fitting curves), which are in principle 
error--free. In other words the differences are {\em intrinsically real} 
and reflect directly the change of the shape introduced by the different
mathematical function. The interpretation of this feature is reported
in the Appendix; from that, we conclude that the observed dispersion is
strictly related to the smallness of the Fourier component involved. 
In our example, the small second Fourier component has
changed its phase value by several tenths of a radian, while for the other
components the change is much smaller. For the same reason we should expect 
an analogous results for $\phi_{41}$, i.e. we should have some dispersion at 
$P \sim 7$ d, where $R_{41}$ is small since another resonance, 
$P_{0}/P_{4}=3$, should be operating there (e.g. Antonello \cite{ant}). 
Indeed this is shown in 
Fig. \ref{diffe3}; note also that the discontinuity of
$\Delta{R_{21}}$ located at 10 d is replaced by that of $\Delta{R_{41}}$ 
at about 7 d. In a certain sense, plots such as those shown in
Figs. \ref{diffe1}, \ref{diffe2} and \ref{diffe3} are better indicators 
of resonance effects than the classical ones, because they are free of
subjective corrections of the phase differences by $\pm{2\pi}$, which could 
be uncertain, mainly for the higher orders. 
Finally, it is possible to note two minima in the lower panel of 
Fig. \ref{diffe1}, one at the resonance center, and the other 
at $\log P \sim 1.5$. Kanbur et al. 
(\cite{kan}) noted the structural change of the light curves at this
$P$; these features still await a theoretical interpretation.

Last but not least, we remark further that several problems with the time 
series analysis of stellar luminosities would be 
simplified by adopting intensity scales instead of magnitude scales. This
statement is not new, of course. Our comment is just further support to
the proposal of abandoning the magnitudes. 
In fact, the blending has no effect on the light curve shape when
we use intensity light curves, and this is an advantage, since one is always
dealing with observed parameters which are affected by errors.

\appendix
\section{Intensity and magnitudes}
Note that the increasing blending produces a light curve, 
expressed in magnitudes, with a shape which is similar
to the shape of the intensity light curve. That is, for very large $L_c$, the 
amplitude becomes very small, and the Fourier parameters become those of the 
intensity--light curve (for the phase differences one has to consider the 
different sign).  The diagrams of 
the differences between intensity-- and magnitude--light curves of the SMC
Cepheids look similar to those of the diagrams shown in Figs. 3, 4 and 5,
but with slightly different ranges of the ordinatae; for example, the range of
$\Delta{R_{21}}$ values would be about $\pm{0.05}$ instead of about
$\pm{0.035}$ as indicated by Fig. 3.

In this Appendix we will use some approximations to understand the effect 
seen near the resonances between pulsation modes of Cepheids, or more 
generally the effect on the smaller Fourier components, given by different 
mathematical descriptions of the light curve. In this context, the 
intensity--light curve could be considered, in a certain sense, as a 
magnitude--light curve for an extremely large blending value.

Let us assume that the {\em intensity} curve is expressed by
\begin{equation}
L = L_o + x = L_o + \sum[A_i\cos(i\omega{t}) + B_i\sin(i\omega{t})],
\end{equation}
where $L_o$ is the mean intensity value, which may include the contribution
from a companion star or blending, and $\omega$ is the pulsation frequency.
The observed light curve can be written as
\begin{equation}
V = -2.5 \log(L) + k_1,
\end{equation}
where $k_1$ is an appropriate constant. By considering the natural logarithm, 
we can write
\begin{equation}
V' = {\rm ln} (L) + k_1' = {\rm ln} (1 + x/L_o) + k_2,
\end{equation}
where $V'=-V/1.0857$ and $k_2 = - k_1/1.0857 + {\rm ln}(L_o)$.
We assume a relatively small amplitude, and expand (A.3) in the series
\begin{equation}
V' = k_2 + x/L_o - (x/L_o)^2/2 + ...
\end{equation}
where the Fourier series of espression (A.1) is introduced, and we assume
for simplicity that $i \le 3$. After some manipulation, we get the 
following expressions for the coefficients of the cosine terms, 
from $i=1$ to $i=6$,
\begin{eqnarray}
A_1/L_o - (A_1A_2 + B_1B_2 + A_2A_3 + B_2B_3)/2L_o^2\\
A_2/L_o - (A_1A_3 + B_1B_3 + A_1^2/2 + B_1^2/2)/2L_o^2\\
A_3/L_o - (A_1A_2 - B_1B_2)/2L_o^2\\
 - (A_1A_3 - B_1B_3 + A_2^2/2 - B_2^2/2)/2L_o^2\\
 - (A_2A_3 - B_2B_3)/2L_o^2\\
 - (A_3^2/2 - B_3^2/2)/2L_o^2,
\end{eqnarray}
respectively.
Six Fourier components are needed instead of just three to describe 
the $V'$ light curve. Analogously for the sine terms we get
\begin{eqnarray}
B_1/L_o - (A_1B_2 - B_1A_2 + A_2B_3 - B_2A_3)/2L_o^2\\
B_2/L_o - (A_1B_1 + A_1B_3 - B_1A_3)/2L_o^2\\
B_3/L_o - (A_1B_2 + B_1A_2)/2L_o^2\\
 - (A_1B_3 + B_1A_3 + A_2B_2)/2L_o^2\\
 - (A_2B_3 + B_2A_3)/2L_o^2\\
 - (A_3B_3)/2L_o^2,
\end{eqnarray}
and the correcting term for the mean value:
\begin{equation}
-(A_1^2 + B_1^2 + A_2^2 + B_2^2 + A_3^2 + B_3^2)/4L_o^2.
\end{equation}
If we had considered a further cubic power of $x$ in the expansion (A.4),
the previous expressions for the coefficients would have included
another correcting term containing cubic power and cross--products of
$A_i$ and $B_i$ multiplied by 1/3$L_o^3$, and the number of Fourier
components would have been nine.

We will assume that the absolute values of the coefficients 
$A_2, B_2$ are much smaller than those of $A_1, B_1$ and $A_3, B_3$, that is, 
the second Fourier component is very small with respect to the first 
and third ones. We note that here we are not dealing with the nonlinear
oscillator problem (e.g. Antonello \cite{ann1}, \cite{ann2}). In the
coefficient of the second cosine and sine term, (A.6) and
(A.12), the first elements, $A_2/L_o$ and $B_2/L_o$ are, according to 
our assumption, small in comparison with the absolute value of the 
correcting terms which contain squares and cross--products of
$A_1, A_3, B_1, B_3$. On the other hand, for the same reason the corrections 
of the coefficients of the first and third cosine and sine terms are
small. In other words, while the first and third Fourier components
are only slightly changed, we must expect a very different second component of 
the Fourier decomposed $V'$ light curve from that of the $L$ light curve.
This conclusion applies, of course, to any value of blending.

\end{document}